\documentclass[12pt]{article}
\usepackage{epsfig}

\newcommand{\postbb}[3]
{\setlength{\epsfxsize}{#3\hsize}
 \centerline{\epsfbox[#1]{#2}}}

\newcommand{\beq}{\begin{equation}}
\newcommand{\eeq}{\end{equation}}
\newcommand{\beqa}{\begin{eqnarray}}
\newcommand{\eeqa}{\end{eqnarray}}
\def\ra{\rightarrow}
\def\RA{\rightarrow}

\newcommand{\siz}{\mbox{$\sin^2\hat{\theta}_W(M_Z)\ $}} 
                                
\newcommand{\mt}{\mbox{$m_t$}}                                                  
\newcommand{\mh}{\mbox{$M_H$}}                                                  
\newcommand{\mz}{\mbox{$M_Z$}}                                                  
                                                  
\newcommand{\alsz}{\mbox{$\alpha_s(M_Z)$}}

\newcommand{\ZP}[3]{{\em Zeit. Phys.} {\bf #1} (19#3) #2}
\newcommand{\MPL}[3]{{\em Mod. Phys. Lett.} {\bf #1} (19#3) #2}
\newcommand{\PR}[3]{{\em Phys. Rev.} {\bf #1} (19#3) #2}
\newcommand{\PL}[3]{{\em Phys. Lett.} {\bf #1}  (19#3) #2}

\newcommand{\PRL}[3]{{\em Phys. Rev. Lett.} {\bf #1}  (19#3) #2}
\newcommand{\NP}[3]{{\em Nucl. Phys.} {\bf #1}  (19#3) #2}

\catcode`\@=11
\long\def\@makefntext#1{
\protect\noindent \hbox to 3.2pt {\hskip-.9pt  
$^{{\ninerm\@thefnmark}}$\hfil}#1\hfill}		

\def\@makefnmark{\hbox to 0pt{$^{\@thefnmark}$\hss}}  
	
\def\ps@myheadings{\let\@mkboth\@gobbletwo
\def\@oddhead{\hbox{}
\rightmark\hfil\ninerm\thepage}   
\def\@oddfoot{}\def\@evenhead{\ninerm\thepage\hfil
\leftmark\hbox{}}\def\@evenfoot{}
\def\sectionmark##1{}\def\subsectionmark##1{}}

\renewcommand{\thefootnote}{\fnsymbol{footnote}}

\newcounter{sectionc}\newcounter{subsectionc}\newcounter{subsubsectionc}
\renewcommand{\section}[1] {\vspace*{0.6cm}\addtocounter{sectionc}{1} 
\setcounter{subsectionc}{0}\setcounter{subsubsectionc}{0}\noindent 
	{\normalsize\bf\thesectionc. #1}\par\vspace*{0.4cm}}
\renewcommand{\subsection}[1] {\vspace*{0.6cm}\addtocounter{subsectionc}{1} 
	\setcounter{subsubsectionc}{0}\noindent 
	{\normalsize\it\thesectionc.\thesubsectionc. #1}\par\vspace*{0.4cm}}
\renewcommand{\subsubsection}[1]
{\vspace*{0.6cm}\addtocounter{subsubsectionc}{1}
	\noindent {\normalsize\rm\thesectionc.\thesubsectionc.\thesubsubsectionc. 
	#1}\par\vspace*{0.4cm}}

\newcounter{appendixc}
\newcounter{subappendixc}[appendixc]
\newcounter{subsubappendixc}[subappendixc]

\renewcommand{\appendix}[1] {\vspace*{0.6cm}
        \refstepcounter{appendixc}
        \setcounter{figure}{0}
        \setcounter{table}{0}
        \setcounter{equation}{0}
        \renewcommand{\thefigure}{\Alph{appendixc}.\arabic{figure}}
        \renewcommand{\thetable}{\Alph{appendixc}.\arabic{table}}
        \renewcommand{\theappendixc}{\Alph{appendixc}}
        \renewcommand{\theequation}{\Alph{appendixc}.\arabic{equation}}
        \noindent{\bf Appendix \theappendixc #1}\par\vspace*{0.4cm}}

\def\abstracts#1{{
	\centering{\begin{minipage}{12.2truecm}\footnotesize\baselineskip=12pt\noindent
	\centerline{\footnotesize ABSTRACT}\vspace*{0.3cm}
	\parindent=0pt #1
	\end{minipage}}\par}} 

\renewenvironment{thebibliography}[1]
	{\begin{list}{\arabic{enumi}.}
	{\usecounter{enumi}\setlength{\parsep}{0pt}
\setlength{\leftmargin 1.25cm}{\rightmargin 0pt}
	 \setlength{\itemsep}{0pt} \settowidth
	{\labelwidth}{#1.}\sloppy}}{\end{list}}

\newcounter{itemlistc}
\newcounter{romanlistc}
\newcounter{alphlistc}
\newcounter{arabiclistc}

\newcommand{\fcaption}[1]{
        \refstepcounter{figure}
        \setbox\@tempboxa = \hbox{\footnotesize Fig.~\thefigure. #1}
        \ifdim \wd\@tempboxa > 6in
           {\begin{center}
        \parbox{6in}{\footnotesize\baselineskip=12pt Fig.~\thefigure. #1}
            \end{center}}
        \else
             {\begin{center}
             {\footnotesize Fig.~\thefigure. #1}
              \end{center}}
        \fi}

\newcommand{\tcaption}[1]{
        \refstepcounter{table}
        \setbox\@tempboxa = \hbox{\footnotesize Table~\thetable. #1}
        \ifdim \wd\@tempboxa > 6in
           {\begin{center}
        \parbox{6in}{\footnotesize\baselineskip=12pt Table~\thetable. #1}
            \end{center}}
        \else
             {\begin{center}
             {\footnotesize Table~\thetable. #1}
              \end{center}}
        \fi}

\def\@citex[#1]#2{\if@filesw\immediate\write\@auxout
	{\string\citation{#2}}\fi
\def\@citea{}\@cite{\@for\@citeb:=#2\do
	{\@citea\def\@citea{,}\@ifundefined
	{b@\@citeb}{{\bf ?}\@warning
	{Citation `\@citeb' on page \thepage \space undefined}}
	{\csname b@\@citeb\endcsname}}}{#1}}

\newif\if@cghi
\def\cite{\@cghitrue\@ifnextchar [{\@tempswatrue
	\@citex}{\@tempswafalse\@citex[]}}
\def\citelow{\@cghifalse\@ifnextchar [{\@tempswatrue
	\@citex}{\@tempswafalse\@citex[]}}
\def\@cite#1#2{{$\null^{#1}$\if@tempswa\typeout
	{IJCGA warning: optional citation argument 
	ignored: `#2'} \fi}}

 1
 1
 1

\font\ninerm=cmr9

\begin{document}

\centerline{\normalsize\bf UNIFICATION OR COMPOSITENESS?\footnote{
Presented at the {\it Ringberg Workshop on the Higgs Puzzle}, 12/96.}}
\baselineskip=16pt

\centerline{\footnotesize PAUL LANGACKER}
\baselineskip=13pt
\centerline{\footnotesize\it Department of Physics and Astronomy,
University of Pennsylvania}
\baselineskip=12pt
\centerline{\footnotesize\it Philadelphia, PA 19104-6396, USA}
\centerline{\footnotesize E-mail: pgl@langacker.hep.upenn.edu}
\vspace*{0.3cm}
\centerline{\footnotesize and}
\vspace*{0.3cm}
\centerline{\footnotesize JENS ERLER}
\baselineskip=13pt
\centerline{\footnotesize\it Institute for Particle Physics, 
University of California, Santa Cruz CA 95064, USA}
{\centerline \footnotesize UPR-0743T \hfill \today}

\vspace*{0.9cm}
\abstracts{The status of precision electroweak data, tests of the
standard model, determination of its parameters, and constraints
on new physics, are surveyed.}
 
\normalsize\baselineskip=15pt
\setcounter{footnote}{0}
\renewcommand{\thefootnote}{\alph{footnote}}
%

\section{Unification or Compositeness?}
Most extensions of the standard model fall into one of two general categories,
unification or compositeness. Unification theories, which include
grand unification and string theories, typically involve a grand desert
between the electroweak and the string or unification scales. They
usually include elementary Higgs fields. The most popular versions
involve supersymmetry, broken at the electroweak or TeV scales, and employ the 
cancellation between ordinary and superpartner contributions to the
Higgs mass renormalization to avoid large radiative corrections to the
electroweak scale. The approximate unification of gauge couplings is perhaps a 
hint that this approach is correct. In that case, the most likely
types of new physics at the TeV scale are generally limited
to superpartners; an extra Higgs doublet;
and possibly additional heavy $Z$ bosons, certain types of
exotic vector multiplets, and 
gauge singlets. In such models, the new physics tends to decouple
from precision observables, i.e., to yield corrections which vanish as the
particle masses become large. In particular, flavor changing neutral currents 
and new sources of CP violation should be small (but not necessarily negligible),
and corrections to precision  experiments such as $Z$ pole measurements are
expected to be very small for most of parameter space.

Another possibility is compositeness - composite fermions and/or
dynamical mechanisms for electroweak symmetry breaking instead of
elementary Higgs fields. Composite quarks or leptons
would not be
analogous to previous levels of compositeness, all of which 
were weakly bound: experimental limits
indicate that any quark or lepton constituent masses should be
at least of the TeV scale, so that
any new level must involve very strong binding. 
Dynamical symmetry breaking models avoid elementary Higgs fields
and therefore avoid naturalness problems associated with quadratic
divergences in the self-energies of elementary Higgs fields.
Models with composite fermions tend to have large flavor changing
neutral current effects due to constituent interchange.
Dynamical symmetry breaking models usually have unacceptable
flavor changing effects due to the exchange of new gauge bosons or
bound states unless the relevant mass scales are very large. Even if 
flavor changing problems are avoided, simple examples of these schemes
usually involve sizable (several per cent) 
contributions to precision observables, either
due to new four-fermion operators or to non-decoupling effects (radiative
corrections that do not vanish as the mass scale increases). Such models
also do not generally predict the apparent gauge unification, although
the latter could conceivably be an accident.

The precision electroweak measurements, especially the LEP and SLD
$Z$ pole observables, have verified the standard model predictions at
the level of a few tenths of a percent. This is consistent with
the expectations of typical unification-type models,
but not with most of the simple compositeness models. 
This may be considered a strong encouragement for
the unification/supersymmetry approach, which reinforces the
hint from gauge unification. However, it is certainly not a proof that
this is the approach followed by nature - that would require the
direct observation of superpartners at colliders. Depending on one's
point of view, the compositeness/dynamical symmetry breaking
route is either disfavored, or at least one is guided to look for versions
which decouple from both flavor changing effects and precision
electroweak observables \cite{dsb}.

In this talk we update our previous analyses \cite{pdg} of the precision data
for testing the standard model, determining its parameters, and
constraining classes of new physics, using the data
presented at the time of the Warsaw Conference.

\section{Recent Data}
\label{recentdata}
The LEP \cite{lep} and SLD \cite{sld} values of the
main $Z$-pole observables are displayed in Table~\ref{tab1},
along with their standard model expectations. 
Along with the $Z$ mass and (partial) widths, 
many asymmetries and polarizations have been observed. The
latter depend on
\beq A^0_f = \frac{2 \bar{g}_{Vf} \; \bar{g}_{Af}}{\bar{g}^2_{Vf}
+ \bar{g}^2_{Af}},  \label{eqn1} \eeq
where $\bar{g}_{V,Af}$ are the vector and axial vector couplings to fermion
$f$. $M_Z$ has been
determined at LEP to the incredible precision (for high energy) experiments
of around 0.002\%. Using $M_Z$ (as well as $\alpha$ and $G_F$) as input
one can predict the other observables (in Table~\ref{tab1} we use
the values of $m_t$, $\alpha_s$, and
$\Delta \alpha_{had}$ obtained from the global fits, and a reasonable
range for $M_H$).

\begin{table}[h] \centering 
\tcaption{$Z$-pole observables from LEP and SLD compared to their standard
model expectations.  The standard model prediction is based on $M_Z$ and
uses the global best fit values for $m_t$, $\alpha_s$, and
$\Delta \alpha_{had}$, with $M_H$ in the
range $60 - 1000$~GeV. August 1996.}
\small
\begin{tabular}{||lcc||}
\hline \hline
Quantity & Value & Standard Model \\ \hline
$M_Z$ (GeV) & $91.1863 \pm 0.0020$ &  input \\
$\Gamma_Z$ (GeV) & $2.4946 \pm 0.0027$ & $2.496 \pm 0.001 \pm
0.001 \pm [0.002] $ \\
$R = \Gamma({\rm had})/\Gamma(\ell \bar{\ell})$ & $20.778 \pm
0.029$ & $20.76 \pm 0.003 \pm 0.001 \pm [0.02]$ \\
$\sigma_{\rm had} = \frac{12 \pi}{M_Z^2} \; \frac{\Gamma(e
\bar{e}) \Gamma({\rm had})}{\Gamma_Z^2} $ 
      & $41.508 \pm 0.056$
& $41.46 \pm 0.002 \pm 0.002 \pm [0.02]$ \\
$R_b = \Gamma(b \bar{b})/ \Gamma({\rm had})$ &$0.2178 \pm 0.0011$
& $0.2156 \pm 0 \pm 0.0002$ \\
$R_c = \Gamma(c\bar{c}) /\Gamma({\rm had})$ & $0.1715 \pm 0.0056$
 & $0.172 \pm 0 \pm 0$ \\
$A^{0\ell}_{FB} = \frac{3}{4} \left( A_{\ell}^0 \right)^2$ &
$0.0174 \pm 0.0010$ & $0.0157 \pm 0.0003 \pm 0.0003$  \\
$A^0_{\tau} \left(P_\tau \right)$ & $0.1401 \pm 0.0067$ & $0.145
\pm 0.001 \pm 0.001$ \\
$A^0_e \left( P_\tau\right)$ & $0.1382 \pm 0.0076$ & $0.145 \pm
0.001 \pm 0.001$ \\
$A^{0b}_{FB} = \frac{3}{4} A^0_e A^0_b$ & $0.0979 \pm 0.0023$ &
$0.101 \pm 0.001 \pm 0.001$ \\
$A^{0c}_{FB} = \frac{3}{4} A^0_e A^0_c$ & $0.0735 \pm 0.0048$ &
$0.072 \pm 0.001 \pm 0.001$ \\
$\bar{s}^2_{\ell} \left(A^{Q}_{FB} \right)$ & $0.2320 \pm
0.0010$ & $0.2318 \pm 0.0002 \pm 0.0001$ \\
$A^0_{\ell} \left(A^0_{LR}, A^0_{e, \mu, \tau} \right)$ \ \ (SLD) 
& $0.1542 \pm 0.0037 $
& $0.145 \pm 0.001 \pm 0.001$ \\
$A^0_b $ \ \ (SLD) & $0.863 \pm 0.049 $ & $0.935 \pm 0 \pm 0$ \\
$A^0_c $ \ \ (SLD) & $0.625 \pm 0.084 $ & $0.667 \pm 0.001 \pm 0.001$ \\
$N_\nu$ & $2.989 \pm 0.012$ & $3$ \\ \hline \hline
\end{tabular}
\label{tab1}
\end{table}

\begin{figure}[h]
\vspace*{13pt}
\leftline{\hfill\vbox{\hrule width 5cm height0.001pt}\hfill}
\postbb{40 220 530 680}{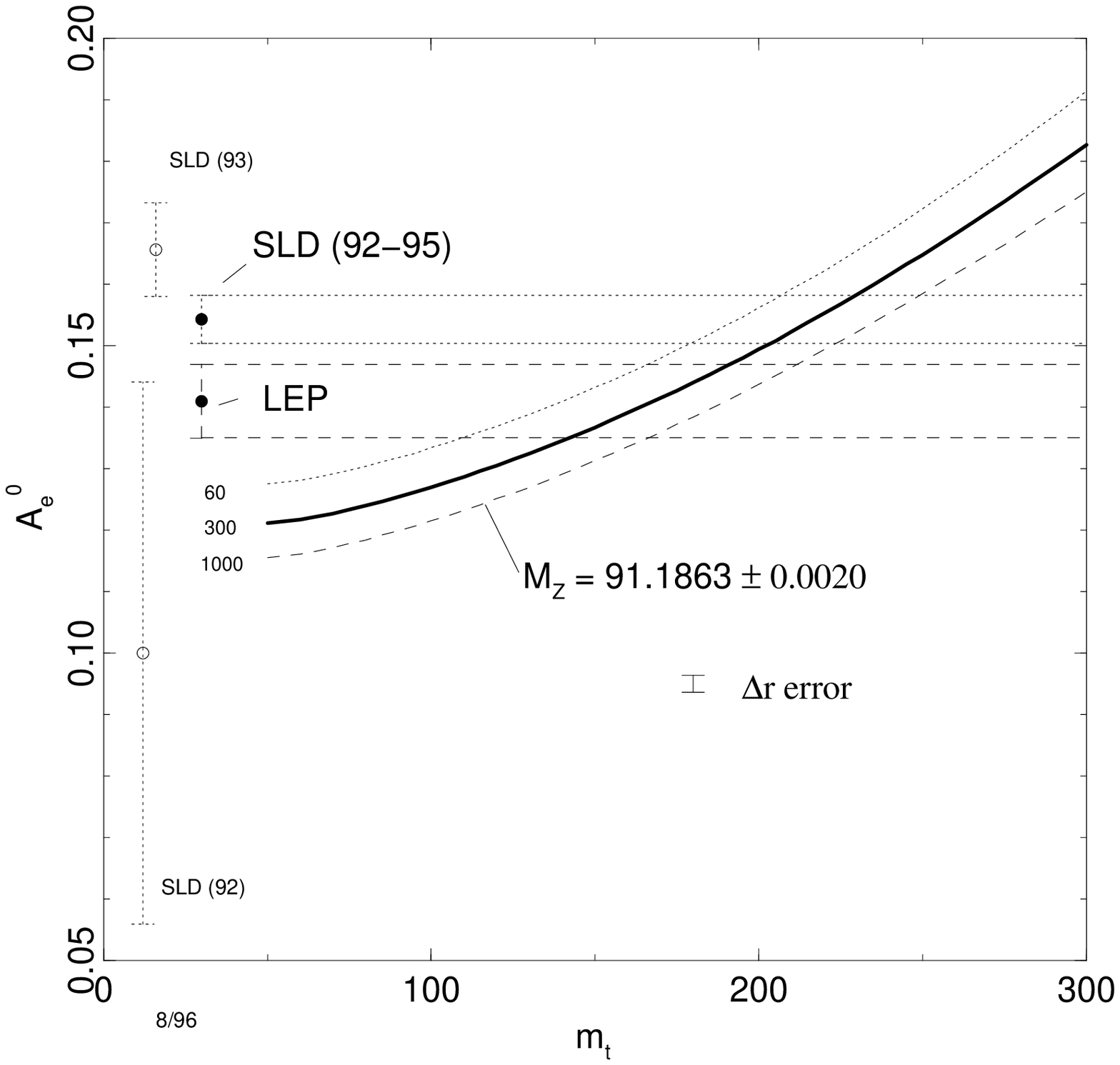}{0.6} 
\leftline{\hfill\vbox{\hrule width 5cm height0.001pt}\hfill}
\fcaption{Values of $A_e$ from SLD and LEP, as well as the 
standard model prediction as a function of $m_t$ for $M_H =$ 60, 300,
and 1000 GeV. 
The direct measurement of $m_t$ from CDF and D0 is $175 \pm 6$ GeV.
 August 1996.}
\label{ae}
\end{figure}

There is generally  impressive agreement between the
standard model predictions and the data. However, there
are two discrepancies at the 2$\sigma$ level. The first is the value of
the leptonic coupling $A^0_\ell \sim 2 \bar{g}_{V\ell}/\bar{g}_{A\ell}$ 
from SLD, which is dominated by the polarization asymmetry $A^0_{LR}
=A^0_e$. The SLD collaboration obtains $A^0_{LR} = 0.1542(37)$,
which is 2.2$\sigma$ above the standard model prediction 0.145(2)
for the allowed $m_t$ range (Figure~\ref{ae}).
This is most likely a fluctuation, because the LEP collaborations
obtain $A^0_e = 0.141(6)$ and $A^0_\ell = 0.147(3)$ from final
state asymmetries/polarizations, in agreement with the standard model.
New, physics, such as a negative $S$ parameter, mixing of the $e_R$
with a heavy exotic lepton, of mixing of the $Z$ with a heavy $Z'$ should
affect both types of observables the same way. The only way to break the
relation would be to have an important contribution to the experiments
that is not directly related to the properties of the $Z$, but it
is difficult to find a sufficiently large mechanism that is not
excluded by other observations \cite{je}. The SLD value for $A^0_{LR}$
(combined with the $Z$ mass) implies $m_t = 217^{+13+20}_{-14-24}$ GeV,
where the second uncertainty is from $M_H$,
well above the direct measurement 175(6) GeV obtained by CDF \cite{cdf}
and D0 \cite{d0}.
Thus, the effect of $A^0_{LR}$ on the global fits is to favor small values
of $M_H$, near the present direct limit of $\sim 65$ GeV.

The other discrepancy is in the ratio $R_b = \Gamma(b \bar{b})/
\Gamma({\rm had})$. The current value, 0.2178(11) is now
2$\sigma$ above the standard model expectation, much closer than
the 3.4$\sigma$ excess reported the year before. The change is mainly due
to new ALEPH results, which are in agreement with the standard model.
($R_c = \Gamma(c \bar{c})/
\Gamma({\rm had})$, which had been $1.8\sigma$ low, is now in agreement.)
The small excess in $R_b$ could still be due to new physics, such as
supersymmetry \cite{susyrb}, mixing with a heavy $Z'$ \cite{zprb}, 
or new extended technicolor\footnote{The simplest ETC models yield large
contributions of the wrong sign.} (ETC) interactions \cite{etcrb}. The effect
is not statistically compelling, but it should be recalled that most
attempts to invoke new physics for the previous larger discrepancy concluded
that it was not possible to obtain $R_b$ larger than 0.218 or to
explain a significant shift in $R_c$, i.e., they predicted precisely
the current values. 

\begin{figure}[h]
\vspace*{13pt}
\leftline{\hfill\vbox{\hrule width 5cm height0.001pt}\hfill}
\postbb{40 220 530 680}{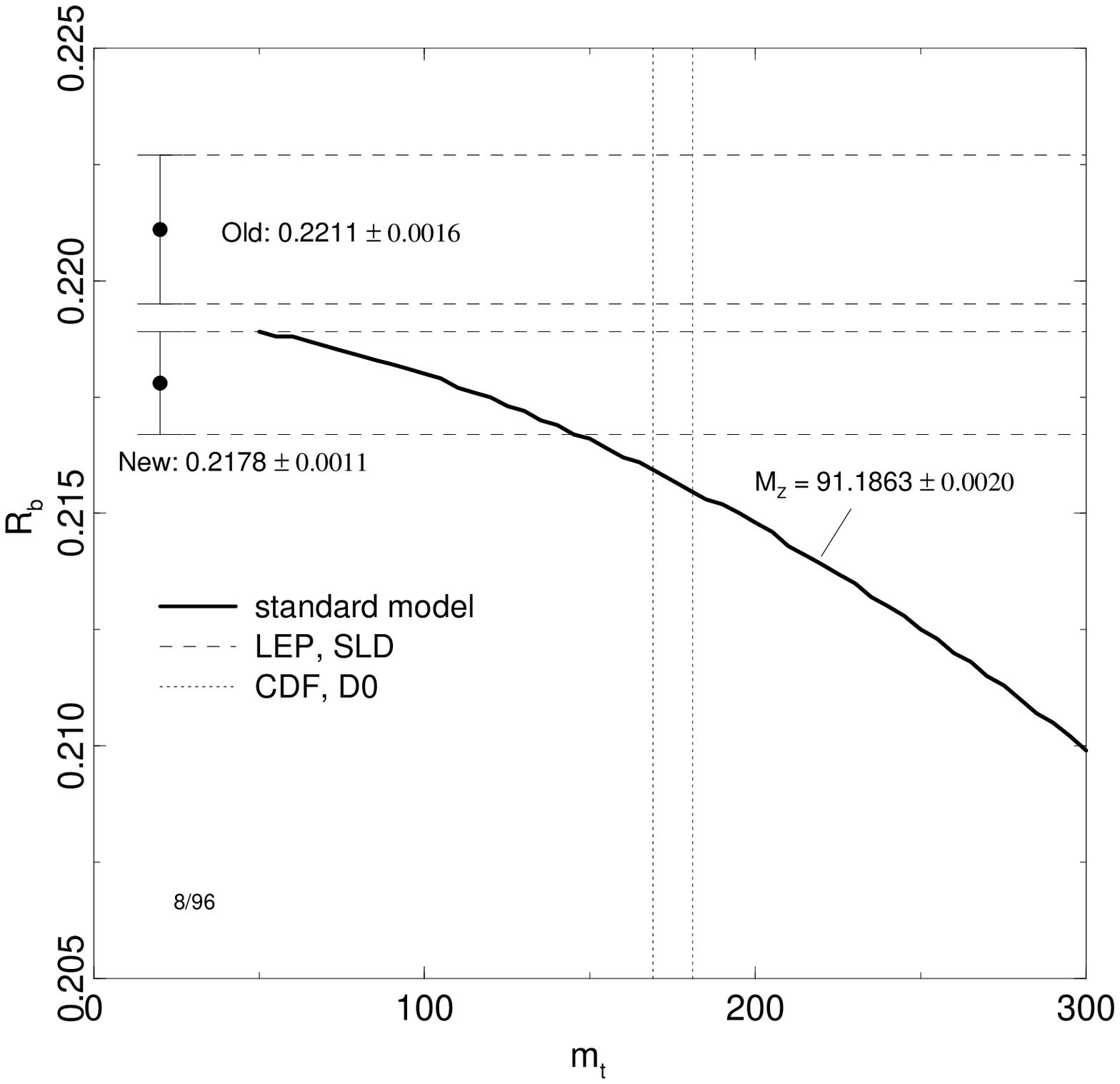}{0.6} 
\leftline{\hfill\vbox{\hrule width 5cm height0.001pt}\hfill}
\fcaption{The current and previous values of $R_b$, as
well as the standard model expectation as a function of $m_t$
and the direct values of $m_t$ from CDF and D0. There is no
significant $M_H$ dependence of the prediction. August 1996.}
\label{rb}
\end{figure}

Nevertheless, the most likely possibility is
a statistical fluctuation. Assuming no new physics, the large
$R_b$ value favors a small $m_t$, as seen in Figure~\ref{rb}.
$R_b$ is by itself insensitive to the Higgs mass, but when combined
with other observables, for which the $m_t$ and $M_H$ dependences
are strongly correlated, $R_b$ favors smaller values of $M_H$.

There is also a strong correlation between $R_b$ and the
strong fine structure constant \alsz. As will be seen in the
Section on \alsz, a precise value of \alsz \ is obtained from
observables related to the hadronic $Z$ width. Any new physics 
contribution to $\Gamma(b \bar{b})$ would imply a 
smaller standard model hadronic width, and therefore a smaller
\alsz. 

There are a number of additional observables, such as the $W$ mass,
results from atomic parity violation, neutral current $\nu-e$ and
$\nu-$ hadron scattering, the direct measurement of the top quark
mass $m_t$ from CDF and D0, and direct limits on the Higgs mass, $M_H$,
from LEP. Some of the more recents results are listed in Table~\ref{tab2}.

\begin{table}[h] \centering 
\tcaption{Recent observables from the $W$ mass and other
non-$Z$-pole observations compared with the standard model
expectations.  Direct limits and values on $M_H$ and $m_t$
are also shown. August 1996.}
\small
\begin{tabular}{||ccc||}  \hline  \hline
Quantity & Value & Standard Model \\ \hline
$M_W$ (GeV) & $80.36 \pm 0.13$ & $80.33 \pm 0.01 \pm 0.03$ \\
$Q_W (C_S)$ & $-71.04 \pm 1.58 \pm [0.88]$ & $-72.85 \pm 0.04 \pm
0.02$ \\
$g_A^{\nu e}$ (CHARM II) & $-0.503 \pm 0.017$ & $-0.506 \pm 0 \pm
0.0002$ \\
$g_V^{\nu e}$ (CHARM II) & $-0.035 \pm 0.017$ & $-0.038 \pm 0.0004
\pm 0.0002$ \\
$s^2_W \equiv 1 - \frac{M_W^2}{M_Z^2}$ & $\begin{array}{c} 0.2213
\pm 0.0048 \;{\rm [CCFR]} \\ 0.2247 \pm 0.0043 \; {\rm [All]}
\end{array}$ & $0.2239 \pm 0.0002\pm 0.0006$ \\
$M_H$ (GeV) & $\geq 65$ LEP 
& $ < O(600), \; {\rm theory}$ \\
$m_t$ & $ 175 \pm 6 $ CDF/D0 & $177 \pm 5 ^{+7}_{-8}$ [with indirect]
\\ \hline \hline
\end{tabular}
\label{tab2}
\end{table}

\section{Fits to the Standard Model and Beyond}
\label{fits}
In the global fits to be described, all of the earlier low energy
observables not listed in Table~\ref{tab2} are fully incorporated.
The electroweak corrections are now quite important.  The results
presented include full
1-loop corrections, as well as  dominant 2-loop effects, QCD corrections,
and mixed QCD-electroweak corrections. 
For the renormalized weak angle, we use
the modified minimal subtraction 
($\overline{MS}$) definition~\cite{msb} 
$\sin^2 \hat{\theta}_W(M_Z) \equiv \hat{s}^2_Z$.  
This basically means that one removes the
$\frac{1}{n-4}$ poles and some associated constants from the gauge
couplings. The fits also include full statistical, systematic,
and theoretical uncertainties, and correlations between the uncertainties.
Our standard model fits are in excellent agreement with those of the LEP 
Electroweak Working Group.

\subsection{The Standard Model and the Decoupled MSSM with Fixed $M_H$}
There are enough independent precision observables to simultaneously
determine \siz, \alsz, and \mt, as well as to constrain additional
parameters such as \mh, the hadronic contribution to the running of
$\alpha$, or parameters representing the effects of new physics. 
In Table~\ref{tab4} we display standard model fits to various
data sets for  fixed values of the Higgs mass. The central
values and first errors in each case are for \mh = 300 GeV, while the
second uncertainties are for $M_H \ra 1000 (+)$ and $60(-)$.
This is a reasonable \mh \ range for the standard model,
including the range between the direct lower bound $\mh > 65$ GeV and
the (rather fuzzy) theoretical upper bound from triviality of
$O(600)$ GeV. (If this were exceeded there would have to be a new physics scale
so close to the Higgs mass to avoid the divergence of the Higgs quartic
coupling that the notion of an elementary Higgs would cease to make sense.)
Also shown in  Table~\ref{tab4} is the Decoupled MSSM fit. This
is the same as the Standard Model fit except that the central value is
\mh $=$ \mz, with a range from $60-150$ GeV. This is relevant to those
versions of the minimal supersymmetric extension of the standard model
 (MSSM) in which the sparticles and second Higgs doublet 
are sufficiently heavy (e.g., a few hundred GeV) that
they decouple, and the only effects of supersymmetry
on the precision observables are that there is a  standard-like
Higgs with mass less than around 150 GeV \cite{susyhiggs}.

\begin{table}[h] \centering
\tcaption{Results for the electroweak parameters in the standard model from
various sets of data.  The central values assume $M_H = 300$~GeV, while the
second errors are for $M_H \ra 1000 (+)$ and $60(-)$.  The last column is
the increase in the overall $\chi^2$ of the fit as $M_H$ increases from 60
to 1000. The last row is for the decoupled MSSM, with a central value
$M_H = M_Z$. The
second errors are for $M_H \ra 150 (+)$ and $60(-)$. August 1996.}
\small
\begin{tabular}{||ccccc||}  \hline \hline
Set & $\hat{s}^2_Z$ & $\alpha_s (M_Z)$ & $m_t$ (GeV) & $\Delta
\chi^2_H$ \\ \hline \hline
\multicolumn{5}{||l||}{Standard Model} \\ \hline
Indirect $+$ CDF $+$ D0   & $0.2316 (2)
( \begin{array}{c} 2  \\ 4 \end{array} )
$ & $0.121
(3)(2)$ & $177\pm 5^{+7}_{-8}      $ & 9.9 \\
All indirect & $0.2315 (2)(1)$ & $0.121 (3)(2)$ & $179 \pm 7
^{+16}_{-19}$ & 7.5 \\
All LEP  & $0.2318 (2)(1)$ & $0.122 (3)(2)$ &
$172 \pm 8^{+17}_{-19}$ & 4.2 \\
$Z$-pole (LEP $+$ SLD) & $0.2315 (2)(1)$ & $0.121 (3)(2)$ &
$178 ^{+7 \;  +17}_{-8 \;  -19}$ & 7.6 \\
SLD $+ \; M_Z$      & $0.2305 (5)(0)$ & ---            &
$217^{+13 \; +20}_{-14 \; -24}$ & \   \\ \hline \hline
\multicolumn{5}{||l||}{Decoupled MSSM} \\ \hline
Indirect $+$ CDF $+$ D0   & $0.2313 (2)(1)$ & $0.119 (3)
( \begin{array}{c} 1  \\ 0 \end{array} )
$ & $171\pm 5 \pm 2      $ &  \\ \hline \hline
\end{tabular}
\label{tab4}
\end{table}

The second row in Table~\ref{tab4} represents the global fit to
all indirect precision data, but not including the direct
CDF/D0 determination $\mt = 175(6)$ GeV. The fit
predicts $\mt = 179 \pm 7
^{+16}_{-19}$ GeV, in remarkable agreement with the CDF/D0 value.
The first row is the global fit, including the direct \mt \ value.
The decoupled MSSM fit in the last row yields parameters
that are slightly shifted due to the lower \mh \ range.

\subsection{The Standard Model or Decoupled MSSM with $M_H$ Free}
Assuming the validity of the standard model one can use the
precision data to constrain the Higgs mass, \mh. Unlike \mt,
which affects the radiative corrections quadratically, the
\mh \ dependence is only logarithmic. Furthermore,
the weaker \mh \ dependence can be comparable to the effects
of new physics, so any constraints or predictions on \mh \  are less robust 
than those on \mt, i.e., they can be modified or lost if there is
any significant contribution from new physics. 
Nevertheless, the current data shows a strong tendency
towards low values of \mh. This can be seen from the last column of
Table~\ref{tab4}, which shows the increase in $\chi^2$ in the best
fit (with respect to the other parameters) as \mh \ is increased from
60 to 1000 GeV. This tendency for a small \mh \
is consistent with the MSSM in the
decoupling limit, which differs from the standard model for the
existing precision data only by the expectation that the (standard-like)
Higgs scalar should be light (less than $\sim$ 150 GeV).
Of course, even if a light Higgs were observed directly at LEP II or
elsewhere it would not by itself prove the existence of supersymmetry,
but it would be extremely encouraging to supersymmetry advocates.

\begin{figure}[h]
\vspace*{13pt}
\leftline{\hfill\vbox{\hrule width 5cm height0.001pt}\hfill}
\postbb{40 220 530 680}{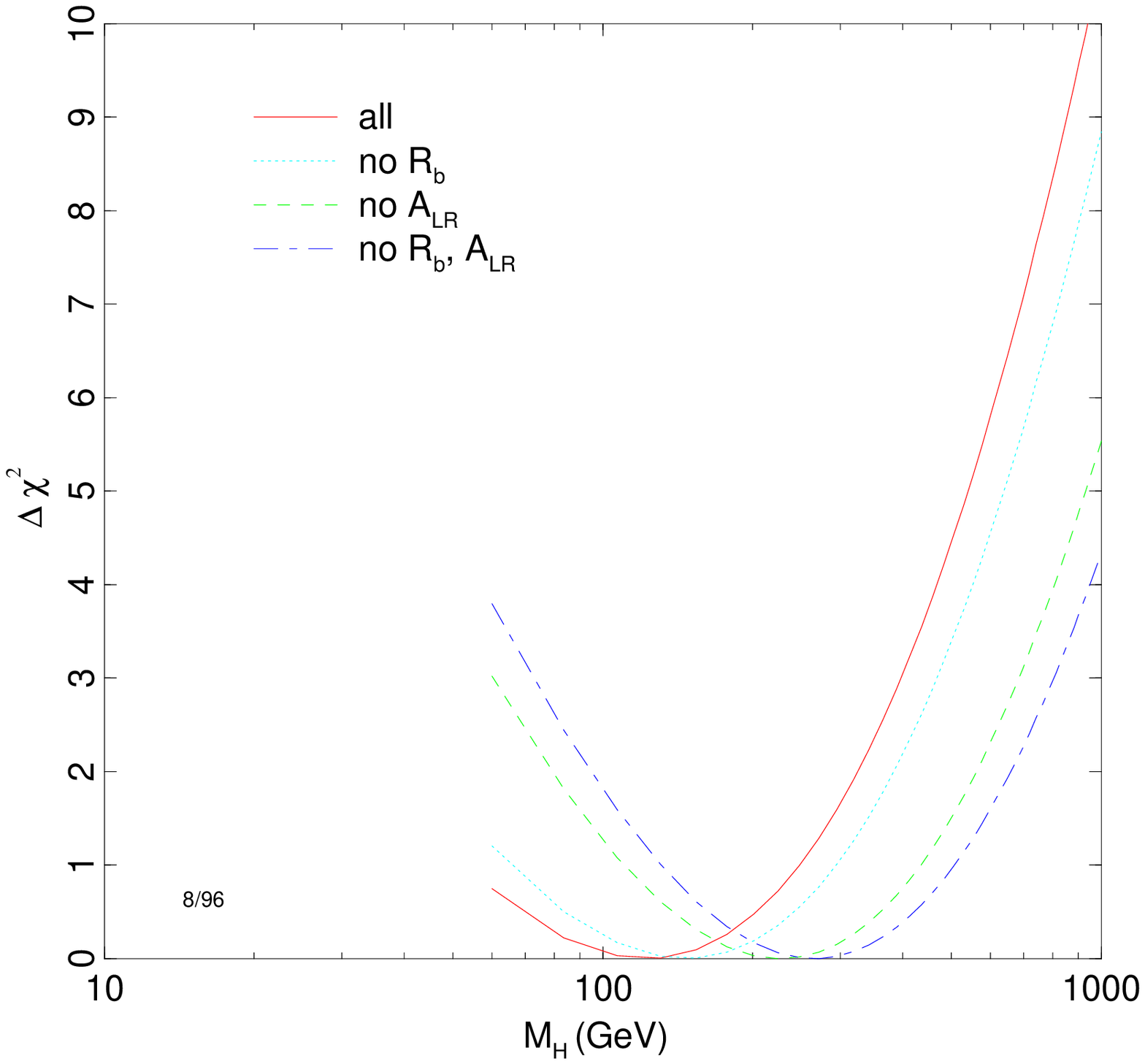}{0.6} 
\leftline{\hfill\vbox{\hrule width 5cm height0.001pt}\hfill}
\fcaption{The increase $\Delta \chi^2$ compared to the best fit as a
function of $M_H$ for various data sets. The corresponding
upper limits (including the direct constraint $M_H > 65$ GeV)
are $M_H <$ 300, 380, and 570 GeV at
90, 95, 99 \% CL. However, the result is driven mainly
by $R_b$ and $A_{LR}$.  August 1996.}
\label{chisq}
\end{figure}

\begin{figure}[h]
\vspace*{13pt}
\leftline{\hfill\vbox{\hrule width 5cm height0.001pt}\hfill}
\postbb{40 220 530 680}{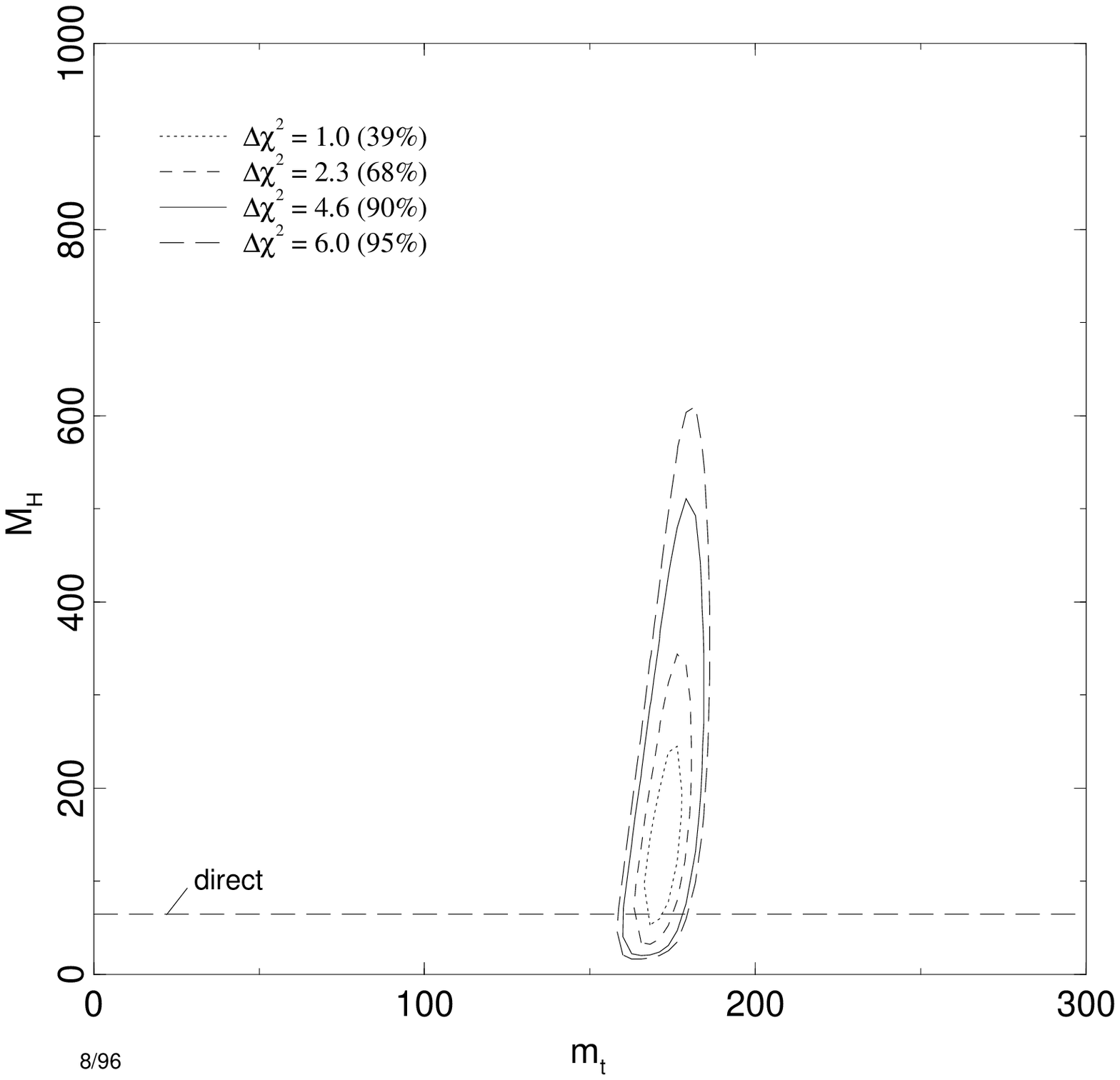}{0.6} 
\leftline{\hfill\vbox{\hrule width 5cm height0.001pt}\hfill}
\fcaption{Allowed regions in $M_H$ and $m_t$ \ at various
confidence levels, including the direct \mt \ constraint. August 1996.}
\label{mh}
\end{figure}

The tendency for a light Higgs is shown in more detail in
Table~\ref{higgsfree} and in Figures~\ref{chisq} and~\ref{mh}.
Leaving \mh \ as a free parameter, one obtains $\mh = 124^{+125}_{-71}$ GeV,
and slightly lower central values for \siz and \alsz \ than in the
$\mh = 300$ GeV fit in Table~\ref{tab4} (but consistent with the
Decoupled MSSM fit). 

The $\chi^2$ distribution as a function of \mh \ is
shown in Figure~\ref{chisq}. The corresponding upper limits on \mh,
which properly take into account the direct lower limit 
$M_H > 65$ GeV and the fact that \mh \ enters the observables
logarithmically, are given in the caption. Some caution is
in order: much of the \mh \ sensitivity and constraint is due to $R_b$
and $A_{LR}$, both of which differ from the
standard model expectation at $\sim 2\sigma$. If there is any new
physics contribution to these quantities, the \mh \ constraint
would be weaked or modified, as is also displayed in Figure~\ref{chisq}.

\begin{table}[h] \centering
\tcaption{Results for the electroweak parameters in the 
standard model, leaving the Higgs mass, $M_H$, free.  The
direct constraint $M_H > 65$ GeV is not included.   August 1996.}
\small
\begin{tabular}{||ccccc||}  \hline \hline
Set & $\hat{s}^2_Z$ & $\alpha_s (M_Z)$ & $m_t$ (GeV) & $M_H$ \\ \hline \hline
Indirect $+$ CDF $+$ D0   & $0.2314 (2)
$ & $0.119
(3)$ &  $172 (6)$ & $124^{+125}_{-71} $   \\ \hline \hline
\end{tabular}
\label{higgsfree}
\end{table}

\subsection{Values of $\alpha_s$ at the $Z$-pole}
\label{alphasvalue}
The hadronic $Z$ width and partial widths receive significant
QCD corrections. Neglecting fermion mass effects (the $m_b$ \ and $m_t$
\ effects are important and are included in the numerical analysis),
\beq
\Gamma(q \bar{q}) = \Gamma^0(q \bar{q}) \left[
1 + \frac{\alsz}{\pi} + 1.409 \left( \frac{\alsz}{\pi} \right)^2
-12.77 \left( \frac{\alsz}{\pi} \right)^3 + {\rm H.O.T.} \right]
\eeq
The $Z$ \ lineshape data  is probably the cleanest determinant
of \alsz \ as far as theoretical QCD uncertainties are involved.
From Table~\ref{tab4} we see that $\alsz=0.121(3)(2)$ for the
Standard Model fit, and $0.119(3)( \begin{array}{c} 1  \\ 0 \end{array} )$
in the Decoupled MSSM\footnote{There is an additional theoretical uncertainty
of $\sim$ 0.001 from higher order terms \cite{ks}.}.
These have come down by $\sim 0.002$ compared
to previous results, and are now consistent with most other
determinations \cite{alphas}, as seen in Table~\ref{tab5}. In particular,
the value $0.118 \pm 0.003$ obtained from lattice calculations of the
$b \bar{b}$ spectrum \cite{lattice} has moved up by 0.003 from the
previously quoted value. The value given in Table~\ref{tab5}
for deep inelastic scattering is still low, but preliminary
recent results from CCFR and BCDMS (not included) are expected to
increase the deep inelastic value.
Thus, most determinations are converging on the value
\alsz $\sim$ 0.118 that has been quoted in the Particle Data Book
\cite{alphas} for some time, and the argument \cite{shifman} for a discrepancy
between the $Z$-lineshape value of \alsz \ and those obtained by
extrapolating low energy data (including QCD sum rule
results not listed in Table~\ref{tab5}) are considerably weakened.

These values of \alsz \ are reasonably consistent with the
prediction $\alsz \sim 0.130 \pm 0.010$ of supersymmetric
gauge coupling unification \cite{gut}, in which the precisely known
$\alpha$ and \siz are used as inputs to predict the unification scale
and \alsz. However, the observed values are on the low side, implying
O(10\%) corrections from threshold effects, non-renormalizable
operators, exotic multiplets, etc. In contrast, the
prediction $\sim$ 0.07 of non-supersymmetric gauge unification
would require much larger corrections.

There is still one uncertainty, however; the lineshape value of \alsz \
is sensitive to any new physics which affects the hadronic width.
In particular, if the 2$\sigma$ excess in $R_b$ is real, and not
just a fluctuation, the extracted \alsz \ would decrease.
This can be quantified by introducing a new physics parameter
$\delta^{new}_{b\bar{b}}$ such that
\beq
\Gamma(b \bar{b}) = \Gamma^{SM}(b \bar{b}) (1 + \delta^{new}_{b\bar{b}}).
\label{eqnbbbar} 
\eeq
$\delta^{new}_{b\bar{b}}$ and \alsz \ are strongly correlated,
and one obtains 
$\alsz = 0.111(5) ( \begin{array}{c} 2  \\ 1 \end{array} )$
in the combined fit, consistent with some low energy values \cite{shifman}.

Thus, if the apparent excess in $R_b$ is due to a fluctuation, the true value of
\alsz \ is most likely around 0.118-0.119, consistent with
supersymmetric gauge unification with moderate theoretical uncertainties.
If the excess is really due to new physics, then a low value for
\alsz \ is called for, requiring large corrections to gauge unification
or abandoning the concept.

\begin{table}[h]  \centering
\tcaption{Values of $\alpha_s$ at the $Z$-pole extracted from
various methods. August 1996.}   
\small                  
\begin{tabular}{||lc||} \hline \hline
Source & $\alpha_s (M_Z)$ \\ \hline
$R_\tau$ & $0.119 \pm 0.004$ \\
Deep inelastic & $0.112 \pm 0.005$ \\
$\Upsilon$, $J/\Psi$ & $ 0.113 \pm 0.006$ \\
$c \bar{c}$ spectrum (lattice) & $0.111 \pm 0.005$ \\
$b \bar{b}$ spectrum (lattice) & $0.118 \pm 0.003$ \\
LEP, lineshape & $0.121 \pm 0.004 $ \\
LEP, event topologies & $0.123 \pm 0.006$ \\ \hline \hline
\end{tabular}
\label{tab5}
\end{table}


\subsection{The Standard Model with $\Delta \alpha_{had}$ Free}
The largest theoretical uncertainty in the standard 
model is the value of $\Delta \alpha_{had}$ from hadronic loops, 
which determines
$\alpha(M_Z)$:
\beq \alpha(M_Z) = 
\frac{\alpha}{1 - \Delta \alpha_{had}- \Delta \alpha_{t}
- \Delta \alpha_{lep}}, \eeq
where $\Delta \alpha_{lep} = 0.03142$ and
$\Delta \alpha_{t} = -0.000061$ represent the leptonic and $t$ quark loops.
$\Delta \alpha_{had}$ can be calculated non-perturbatively
from a dispersion integral over experimental low energy $e^+ e^- \RA$ hadrons
data.
There has been considerable recent work reevaluating $\Delta \alpha_{had}$,
with the results in Table~\ref{hadloops} now in reasonable agreement with
each other. In most of our fits, we use the value 0.0280(7)
of Eidelman and Jegerlehner~\cite{ej}.

Despite the agreement, $\Delta \alpha_{had}$ is still a significant
uncertainty (ten times more important than the experimental error
in \mz). A closely related effect dominates the uncertainty in the
theoretical prediction for the muon anomalous magnetic moment, which
will be considerably larger than the projected experimental error
from the new Brookhaven $g_\mu = 2 $ experiment unless new measurements
are made of the low energy cross section for $e^+ e^- \RA$ hadrons.

It is amusing that the precision data itself can constrain $\Delta \alpha_{had}$,
which enters in the relation between \mz \ and \siz, since \siz is
independently constrained by the asymmetry measurements, and \mt \ is
measured directly \cite{stuart,alhadpaper}. The values for 
$\Delta \alpha_{had}$ and the corresponding $\alpha(M_Z)$ are
shown in Table~\ref{hadloops} for both the standard model and
constrained MSSM Higgs mass ranges. It is seen that for fixed \mh \
the precision is comparable to the independent estimates. For the
standard model case, the uncertainty from \mh \ in the range 60-1000 GeV
is considerably larger, while for the constrained MSSM the Higgs uncertainty
is reasonably small. It is remarkable that the precision data are so good
as to allow the extraction of $\Delta \alpha_{had}$ simultaneously with
the other parameters (such as \siz \ and \alsz). The values
of $\Delta \alpha_{had}$ using both the precision data and the
independent Eidelman and Jegerlehner value are also listed in
Table~\ref{hadloops}.

\begin{table}[h] \centering
\tcaption{ It is now possible to determine
$\Delta \alpha_{had}$ directly from the precision data, with a value 
comparable to independent theoretical estimates \cite{ej}-\cite{swartz}
using low energy
$e^+ e^-$ data. Other fits include Eidelman and Jegerlehner (95)
as a separate constraint. August 1996.}
\small
\begin{tabular}{||lll||}  \hline \hline
Source & $\Delta \alpha_{had}$ & $\alpha(M_Z)$  \\ \hline \hline
Eidelman, Jegerlehner (95) & 0.0280(7) & 128.90 (9) \\
Martin, Zeppenfeld (95) & 0.0273(4) & 128.99 (5) \\
Burkhardt, Pietrzyk (95) & 0.0280(7) & 128.89 (9) \\
Swartz (95) & 0.0275(5) & 128.96 (6) \\ \hline
SM fit, including EJ (95) & 0.0274(5)$(\begin{array}{c} -6 \\ +9 \end{array}) $
& 128.98 (7)$(\begin{array}{c} +8 \\ -12 \end{array}) $ \\
MSSM fit, including EJ (95) & 0.0281(5)$(\begin{array}{c} -3 \\ +2 \end{array}) $
& 128.89 (7)$(\begin{array}{c} +4 \\ -3 \end{array})$  \\ \hline \hline
unconstrained SM fit & 0.0265(9)$(\begin{array}{c} -18 \\ +23 \end{array})$
 & 129.11 (12) $(\begin{array}{c} +25 \\ -32 \end{array})$
  \\
unconstrained MSSM fit & 0.0282(9)$(\begin{array}{c} -7 \\ +6 \end{array})
$ & 128.87 (12) $(\begin{array}{c} +10 \\ -8 \end{array})$
  \\ \hline \hline
\end{tabular}
\label{hadloops}
\end{table}


\section{Beyond the Standard Model}
There are many types of new physics that are constrained by the 
precision data, including new contact operators, heavy $Z'$ bosons,
and mixing between ordinary and exotic fermions. Here we briefly
state the current results for a few parametrizations of certain
classes of new physics. More detailed discussions, as well as
model independent analyses, more discussion of gauge coupling
unification, etc., may be found in~\cite{pdg}.

\subsection{The Standard Model or Decoupled MSSM 
with a $Z b \bar{b}$ Vertex Correction}
The apparent excess in $R_b$ has already been discussed in the Sections
on the data and on \alsz. If one introduces a new
physics parameter $\delta^{new}_{b\bar{b}}$, as in (\ref{eqnbbbar}),
then the other extracted standard model parameters are modified somewhat, as
can be seen in Table~\ref{bbbarfits}.

\begin{table}[h] \centering
\tcaption{
 One can parametrize possible new physics in the  $Z b \bar{b}$ vertex by
$\Gamma(b \bar{b}) = \Gamma^{SM}(b \bar{b}) (1 + \delta^{new}_{b\bar{b}})$.
Allowing $\delta^{new}_{b\bar{b}} \ne 0$ leads to a lower value of $\alpha_s$
extracted from the lineshape. August 1996. }
\small
\begin{tabular}{||ccccc||}  \hline \hline
Set & $\hat{s}^2_Z$ & $\alpha_s (M_Z)$ & $m_t$ (GeV) & 
$\delta^{new}_{b\bar{b}} $ \\ \hline \hline
Indirect $+$ CDF $+$ D0   & $0.2316 (2)
( \begin{array}{c} 2  \\ 4 \end{array} )
$ & $0.121
(3)(2)$ & $177\pm 5^{+7}_{-8}      $ & fixed at 0 \\
 \hline
Indirect $+$ CDF $+$ D0   & $0.2315 (2) (3)
$ & $0.111
(5) ( \begin{array}{c} 2  \\ 1 \end{array} )
$ & $178\pm 5^{+7}_{-8}      $ & $0.014 (7) (2) $ \\
 \hline \hline
\end{tabular}
\label{bbbarfits}
\end{table}

\begin{figure}[h]
\vspace*{13pt}
\leftline{\hfill\vbox{\hrule width 5cm height0.001pt}\hfill}
\postbb{40 220 530 680}{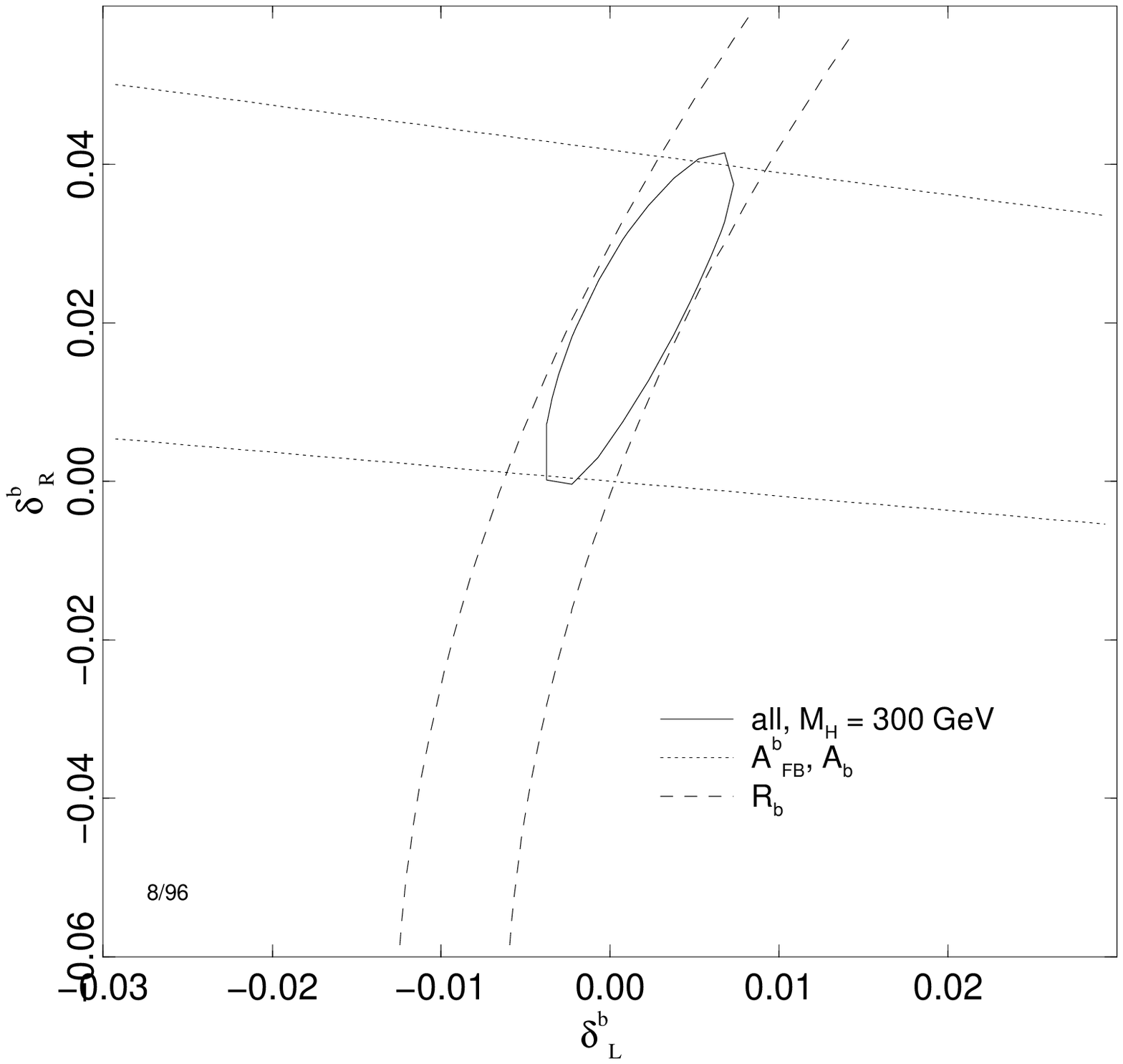}{0.6} 
\leftline{\hfill\vbox{\hrule width 5cm height0.001pt}\hfill}
\fcaption{90\% CL allowed region in
$\delta^b_L$ vs $\delta^b_R$. August 1996.}
\label{blbr}
\end{figure}

A more detailed analysis allows separate corrections to the
left and right chiral $Z b \bar{b}$ vertices, $\delta^b_L$ and
$\delta^b_R$, i.e.
\begin{eqnarray}
g^b_L & \simeq & -\frac{1}{2} + \frac{1}{3} s^2_W + \delta^b_L \sim -0.42 
 + \delta^b_L \nonumber \\ 
 g^b_R  & \simeq & \frac{1}{3} s^2_W + \delta^b_R \sim 0.077 + \delta^b_R.
\end{eqnarray}
A global fit yields
$ \delta^b_L = 0.002(3)(2)$,
$\delta^b_R = 0.02(1)(1)$, and
$\alpha_s(M_Z)= 0.111(5)(1)$.
Thus, the data now favor an anomaly, if any, in $\delta^b_R$, and the
correlation with \alsz \ is essentially unchanged with respect to the single
new parameter case. The allowed region in $\delta^b_L$ vs $\delta^b_R$
is shown in Figure~\ref{blbr}.


\subsection{The $\rho_0$ Parameter}
One parameterization of certain new types of physics is the parameter
$\rho_0$, which is introduced to describe new sources of $SU_2$ breaking
other than the ordinary Higgs doublets or the top/bottom splitting.
One defines $\rho_0 \equiv M_W^2/(M_Z^2 \hat{c}^2_Z \hat{\rho})$,
where $\hat{c}^2_Z \equiv 1 - \hat{s}^2_Z$;
$\hat{\rho} \sim 1 + 3 G_F m_t^2 /8 \sqrt{2} \pi^2 $ 
absorbs the relevant standard model radiative
corrections so that $\rho_0 \equiv 1$ in the standard model.
New physics can affect $\rho_0$ at either the tree or loop-level,
$\rho_0 = \rho_0^{\rm tree} + \rho_0^{\rm loop}$.
The tree-level contribution is given by Higgs representations
larger than doublets, namely,
\beq \rho_0^{\rm tree} = \frac{\sum_i \left( t^2_i -  t_{3i}^2
+ t_i \right) |\langle \phi_i \rangle|^2}{ \sum_i 2 t_{3i}^2
|\langle \phi_i \rangle|^2},  \label{eqerica}\eeq
where $t_i$ ($t_{3i}$) is the weak isospin (third component) of the
neutral Higgs field $\phi_i$.
For Higgs singlets and doublets ($t_i = 0,\frac{1}{2}$) only,
$\rho_0^{\rm tree} = 1$.  
However, $\rho_0^{\rm tree}$ can differ from unity
in the presence of larger representations with non-zero
vacuum expectation values.

One can also have loop-induced contributions similar to that of the
top/bottom, due to non-degenerate multiplets of fermions or bosons.  For new
doublets
\beq \rho_0^{\rm loop} = \frac{3G_f}{8 \sqrt{2} \pi^2} \sum_i
\frac{C_i}{3} F (m_{1i},m_{2i}),   \eeq
where $C_i = 3(1)$ for color triplets (singlets) and
\beq F(m_1, m_2)   =  m_1^2 + m^2_2 - \frac{4m_1^2 \; m^2_2}{m_1^2 -
m^2_2} \ln \frac{m_1}{m_2}  \geq   (m_1- m_2)^2 . \eeq
Loop contributions to $\rho_0$ are generally positive,\footnote{One can
have $\rho^{\rm loop} < 0$ for Majorana fermions 
 or
boson multiplets with vacuum expectation values. 
} and
if present would lead to lower values for the predicted $m_t$. $\rho_0^{\rm
tree} - 1$ can be either positive or negative depending on the quantum numbers
of the Higgs field.  The $\rho_0$ parameter is extremely important because
one expects $\rho_0 \sim 1$ in most superstring theories, which generally
do not have higher-dimensional Higgs representations, while typically
$\rho_0 \neq 1$ from many sources in models involving compositeness.  

\begin{figure}[h]
\vspace*{13pt}
\leftline{\hfill\vbox{\hrule width 5cm height0.001pt}\hfill}
\postbb{40 220 530 680}{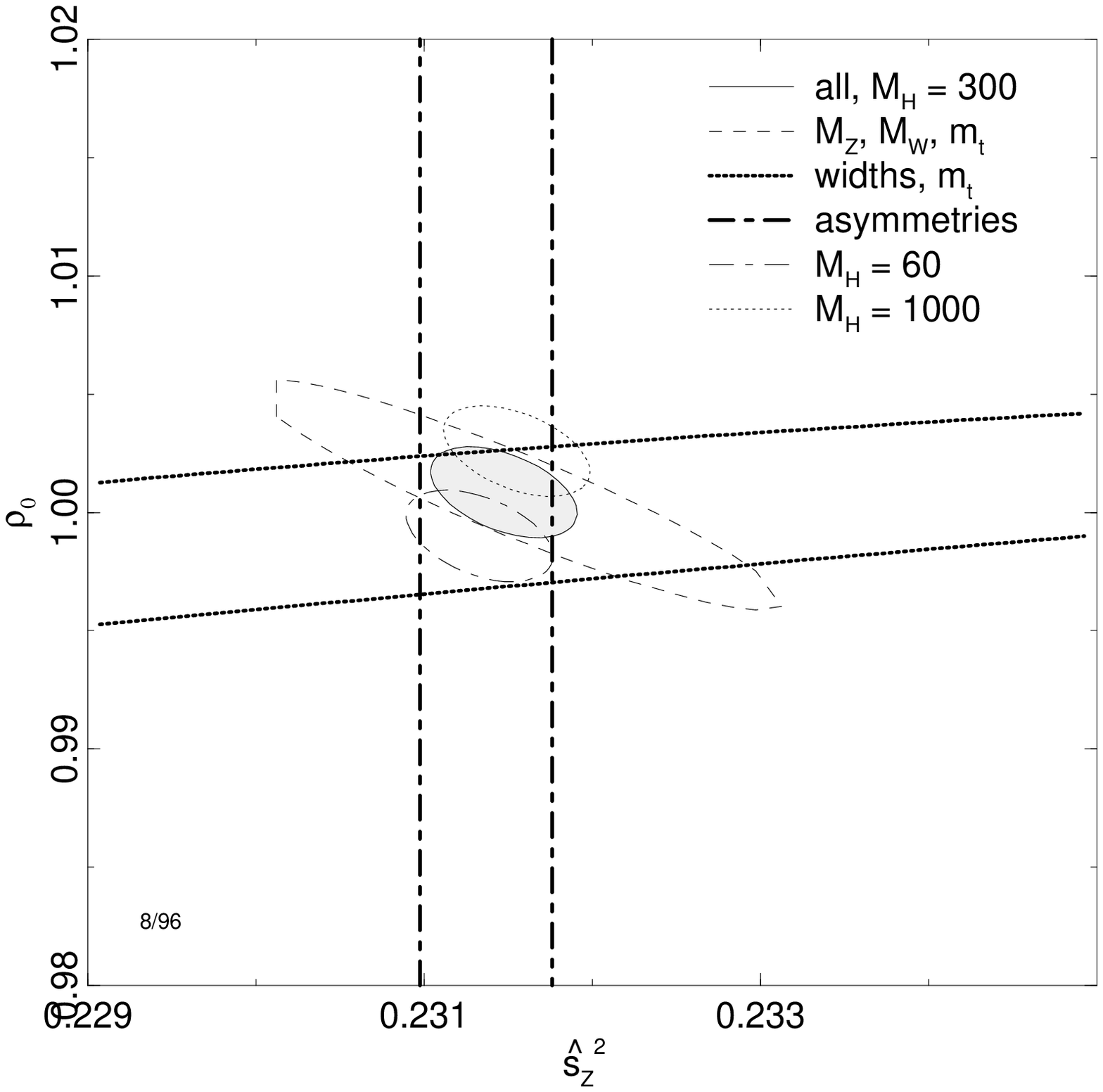}{0.6} 
\leftline{\hfill\vbox{\hrule width 5cm height0.001pt}\hfill}
\fcaption{90\% CL allowed region in $\rho_0$ vs $\siz$. August 1996.}
\label{rho}
\end{figure}

It has long been known that $\rho_0$ is close to 1. However, until recently it
has been difficult to separate $\rho_0$ from $m_t$, because in most
observables one has only the combination $\rho_0 \hat{\rho}$.  The one
exception has been the $Z \ra b\bar{b}$ vertex.  However, the direct 
measurement of \mt \ by CDF and D0 allows one to
calculate $\hat{\rho}$ and therefore separate $\rho_0$.  In practice one
fits to $m_t$, $\rho_0$ and the other parameters, using the CDF/D0 value 
of $m_t$ as an additional constraint.  One can determine
$\hat{s}^2_Z$, $\rho_0$, $m_t$, and $\alpha_s$ simultaneously, yielding
the results listed in Table~\ref{tabrho}.
\begin{table}[h] \centering
\tcaption{
One can parametrize new sources of vector $SU_2$ breaking, such as nondegenerate
new fermion or scalar multiplets, or higher dimensional Higgs multiplets,
by a parameter $\rho_0$, which is exactly unity in the standard
model. Allowing $\delta^{new}_{b\bar{b}} \ne 0$ as well,
one obtains $\rho_0 = 1.0006 (9) (18)$, $\alpha_s = 0.111 (6) (1)$,
$\delta^{new}_{b\bar{b}} = 0.013 (7)$,
with negligible change in the other parameters. August 1996.}
\small
\begin{tabular}{||ccccc||}  \hline \hline
Set & $\hat{s}^2_Z$ & $\alpha_s (M_Z)$ & $m_t$ (GeV) & 
$\rho_0$ \\ \hline \hline
Indirect $+$ CDF $+$ D0   & $0.2316 (2)
( \begin{array}{c} 2  \\ 4 \end{array} )
$ & $0.121
(3)(2)$ & $177\pm 5^{+7}_{-8}      $ & fixed at 1 \\
 \hline
Indirect $+$ CDF $+$ D0   & $0.2315 (2) 
( \begin{array}{c} 1  \\ 2 \end{array} )
$ & $0.119 (4) (1)

$ & $173 (6)     $ & $1.0009 (9) (18) $ \\
 \hline \hline
\end{tabular}
\label{tabrho}
\end{table}
 Even in the presence of the
classes of new physics parameterized by $\rho_0$ one still has robust
predictions for the weak angle and a good determination of $\alpha_s$.
Most remarkably, given the CDF/D0 constraint, $\rho_0$ is constrained to be
very close to unity, causing serious problems
for compositeness models.  The allowed region in $\rho_0$ vs $\hat{s}^2_Z$ are
shown in Figure \ref{rho}.  This places limits $|\langle \phi_i
\rangle|/ |\langle \phi_{1/2} \rangle| < {\rm few} \ \%$ on non-doublet
vacuum expectation values, and places constraints $\frac{C}{3} F(m_1, m_2)
\leq (100\; {\rm GeV})^2$ on the splittings of additional fermion or boson
multiplets. 


\subsection{The $S_{\rm new}$, $T_{\rm new}$, and $U_{\rm new}$ Parameters}
\begin{figure}
\vspace*{13pt}
\leftline{\hfill\vbox{\hrule width 5cm height0.001pt}\hfill}
\postbb{40 220 530 680}{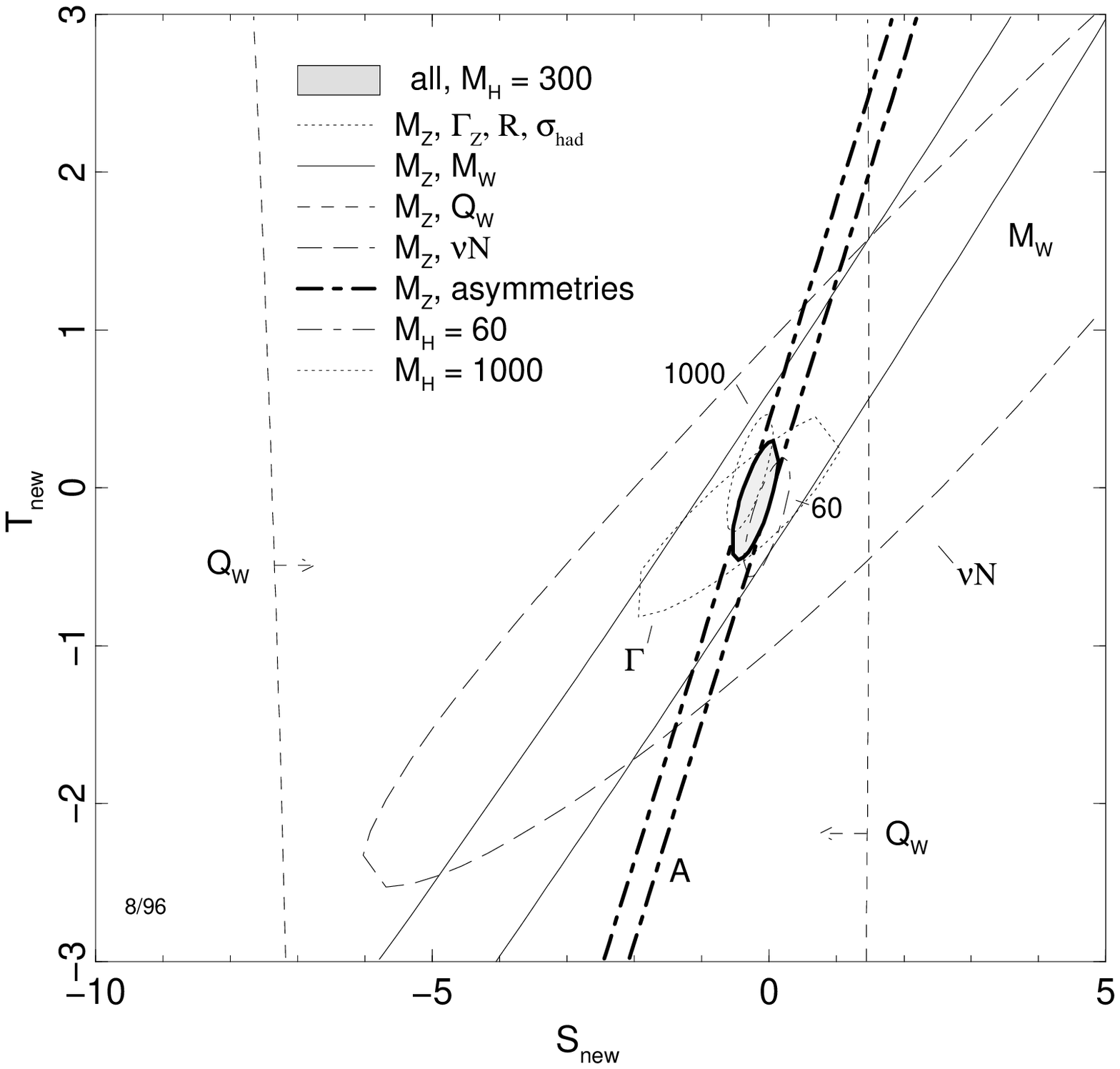}{0.6} 
\leftline{\hfill\vbox{\hrule width 5cm height0.001pt}\hfill}
\fcaption{90\% CL allowed region in $S_{\rm new}$ vs $T_{\rm new}$. August 1996.}
\label{st}
\end{figure}

$S_{\rm new}$, $T_{\rm new}$, and $U_{\rm new}$ 
generalize the $\rho_0$ parametrization of new
physics \cite{sturef}. $S_{\rm new}$ represents new sources of axial $SU_2$ breaking,
such as degenerate chiral multiplets, $T_{\rm new} = (\rho_0 - 1)/\alpha$
represents vector $SU_2$ breaking, including both tree level and
loop effects, while $U_{\rm new}$, which affects
$M_W$, is small in most models. The $S_{\rm new}$, $T_{\rm new}$, 
and $U_{\rm new}$ presented
here are due to new physics only ($m_t$ and $M_H$ effects are
treated separately), and they have a factor of $\alpha$ removed
so that deviations from new physics are expected to be of order
unity. The expectations for these parameters for various types of new
physics and their relation to other equivalent parametrizations are
given in \cite{pdg}. The current values of 
$S_{\rm new}$, $T_{\rm new}$, and $U_{\rm new}$ 
and the standard model parameters are given in Table~\ref{stu}.
The allowed regions in $S_{\rm new}$ and $T_{\rm new}$
are shown in Figure~\ref{st}.
\begin{table}[h] \centering
\tcaption{
Current values of $S_{\rm new}$, $T_{\rm new}$, and $U_{\rm new}$.
 Fits are shown with and without $\delta^{new}_{b\bar{b}}$,
and for the equivalent $\rho_0$ and $\epsilon_i$ parameters \cite{pdg}. The 
standard model (SM) parameters are also shown. August 1996.}
\small
\begin{tabular}{||l|c|c|c||}  \hline \hline
Parameter & SM & $\delta^{new}_{b\bar{b}}=0$ & $\delta^{new}_{b\bar{b}}$ free
\\ \hline \hline
$\hat{s}^2_Z$ & $0.2316 (2)
( \begin{array}{c} 2  \\ 4 \end{array} ) $
&  $0.2313 (2)
( \begin{array}{c} 1  \\ 0 \end{array} ) $ & 0.2313 (2) \\
$\alpha_s (M_Z)$  & $0.121(3)(2)$
&  $0.121 (4) ( \begin{array}{c} 0  \\ 1 \end{array} ) $
  & 0.112 (6) \\
$m_t$ (GeV)  & $177\pm 5^{+7}_{-8}      $ & 173 (6)  & 175 (6) \\
\hline
$S_{\rm new}$  & & $-0.18 (16) ( \begin{array}{c} -8  \\ +17 \end{array} ) $ 
& $-0.19 (16) ( \begin{array}{c} -8  \\ +17 \end{array} ) $ \\
$T_{\rm new}$ &  & $-0.04 (20) ( \begin{array}{c} 17  \\ 11 \end{array} ) $ &
 $-0.08 (19) ( \begin{array}{c} 17  \\ 11 \end{array} ) $ \\
$U_{\rm new}$ &  & 0.07 (42) &  0.06 (42) \\
$\delta^{new}_{b\bar{b}}$ &  & fixed at 0 & 0.013 (7) \\
\hline
$\rho_0$ &  & $0.9997 (15) ( \begin{array}{c} 12  \\ 18 \end{array} ) $
& $0.9994 (14) ( \begin{array}{c} 12  \\ 18 \end{array} ) $ \\
\hline
$\epsilon_3$  & & $-0.0014 (13) ( \begin{array}{c} -7 \\ +13 \end{array} ) $ 
  & $-0.0015 (13) ( \begin{array}{c} -7 \\ +13 \end{array} ) $  \\
$\epsilon_1$  &  & $-0.0003 (14) ( \begin{array}{c} 13 \\ 8 \end{array} ) $ 
& $-0.0006 (13) ( \begin{array}{c} 12 \\ 8 \end{array} ) $  \\
$\epsilon_2$ & & $-0.0005 (33) $ & $-0.0005 (33) $ \\
 \hline \hline
\end{tabular}
\label{stu}
\end{table}

\end{document}